\documentclass{article}
\usepackage{spconf,amsmath,graphicx}
\usepackage{graphicx}

\usepackage{multirow}

\title{WaDeNet: Wavelet Decomposition based CNN for Speech Processing}
\begin{document}
%
\name{Prithvi Suresh$^{\dagger\star}$ \qquad Abhijith Ragav$^{\dagger\star}$
\thanks{$^{\star}$Authors with equal contribution}}
\address{$^{\dagger}$Solarillion Foundation\\
\texttt{\{prithvisuresh, abhijithragav\}@ieee.org}}
\maketitle
\begin{abstract}
  Existing speech processing systems consist of different modules, individually optimized for a specific task such as acoustic modelling or feature extraction. In addition to not assuring optimality of the system, the disjoint nature of current speech processing systems make them unsuitable for ubiquitous health applications. We propose WaDeNet, an end-to-end model for mobile speech processing. In order to incorporate spectral features, WaDeNet embeds wavelet decomposition of the speech signal within the architecture. This allows WaDeNet to learn from spectral features in an end-to-end manner, thus alleviating the need for feature extraction and successive modules that are currently present in speech processing systems. WaDeNet outperforms the current state of the art in datasets that involve speech for mobile health applications such as non-invasive emotion recognition. WaDeNet achieves an average increase in accuracy of 6.36\% when compared to the existing state of the art models. Additionally, WaDeNet is considerably lighter than a simple CNNs with a similar architecture. 
\end{abstract}
\section{Introduction}
\label{sec:intro}
A substantial part of current speech modelling systems is feature extraction \cite{o1988linear,hermansky1985perceptually}. However, the advent of Deep Learning (DL) has caused a significant paradigm shift in how these signals are modelled. Over the last 3 decades, advanced ``hand-crafted'' transformations such as the Mel-frequency cepstral coefficients (MFCC) were used owing to a rise in accuracy in their use with Hidden Markov Model (HMM) systems \cite{muda2010voice}. However, these transformations caused a loss of information. DL models sought to make up for this loss by learning naive spectral \cite{hermansky2011speech} or waveform \cite{sheikhzadeh1994waveform} features from the speech signal thus reducing the dependency on  hand-crafted features. In fact, Deng \textit{et al.} \cite{deng2010binary} showed that DL models benefit from using simpler spectrograms over manually extracted MFCCs. Additionally, when compared to complex features, raw spectral features preserve more information, making them suitable for handling large variability across users (speaking styles, accents, etc.).  

Current systems for speech processing contain different modules that perform different tasks, such as an acoustic module and a classifier \cite{wang2019overview}. Every module has its own training process that optimizes a different objective function, each differing from the true evaluation criteria. Thus, global optimality of the entire system is not guaranteed in spite of module optimality  \cite{zhang2017towards}. Additionally, development of these systems require extensive hyper parameter tuning by experts \cite{miao2015eesen}. These shortcomings curb the ability of current systems to be deployed in a mobile environment, ushering in the need for a powerful end-to-end model.   

Keeping in mind these disadvantages, we propose an end-to-end Deep Learning model for speech processing. Due to its end-to-end nature, these models find use in mobile health applications such as outpatient telemetry and ubiquitous affective computing tasks. We chose to perform our experiments on tasks involving speech towards emotion recognition and pervasive pathological voice detection, since they are important mobile affective computing tasks. The nature of this task makes it appropriate for our experiments since it demands the ability of the solution to be run on edge devices.
\begin{figure*}
    \centering
    \includegraphics[width = \textwidth]{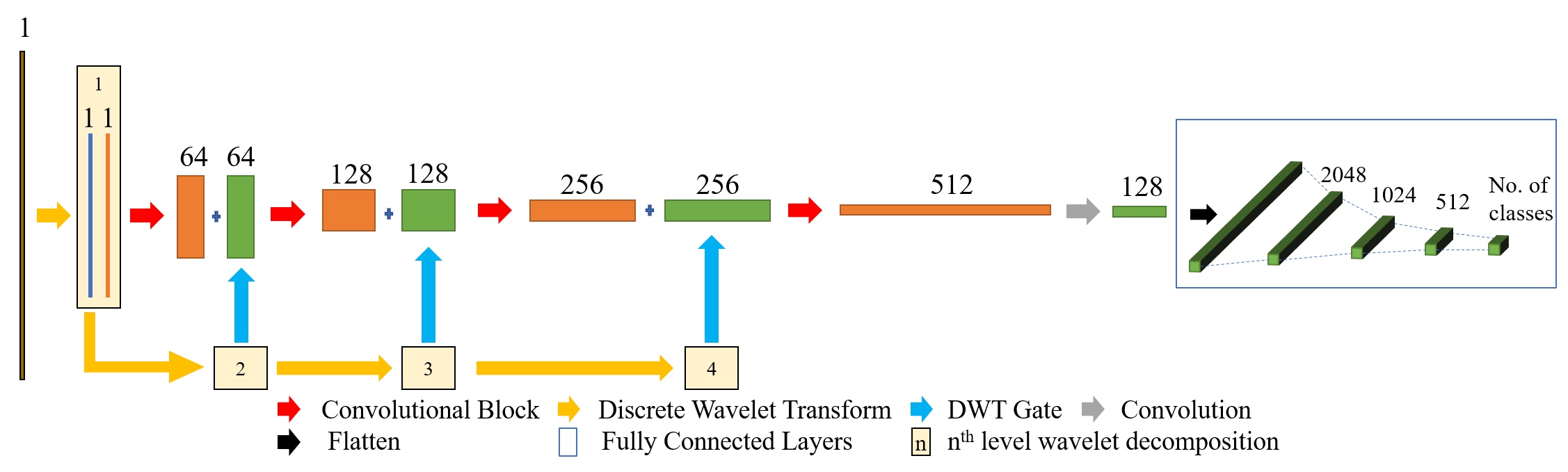}
    \caption{WaDeNet Architecture for $N=4$ ($k=3$)}
    \label{arch}
\end{figure*} 
\section{Methodology}
In an attempt to provide a complete end-to-end solution, an architecture that learns to derive features and suitably learn to classify from these features is to be realised. To this end, a convolutional feature extractor followed by a series of fully connected layers for classification is proposed. Convolutional Neural Network (CNN) architectures have shown success as temporal feature extractors in speech processing tasks \cite{6288864}. By connecting the feature extractor and fully connected layers, and training the resultant neural network, an end-to-end pipeline is achieved. This eliminates the need for feature extraction via complex signal processing techniques. 

For our experiments we used two architectures – a naive 1D CNN (Naive CNN) and an enhanced CNN incorporating wavelet decomposition. Both CNNs contain the same fully connected layers but vary in the feature extractor. Their architectures are explained in succeeding sections. 

\begin{figure}
    \centering
    \includegraphics[width = \columnwidth]{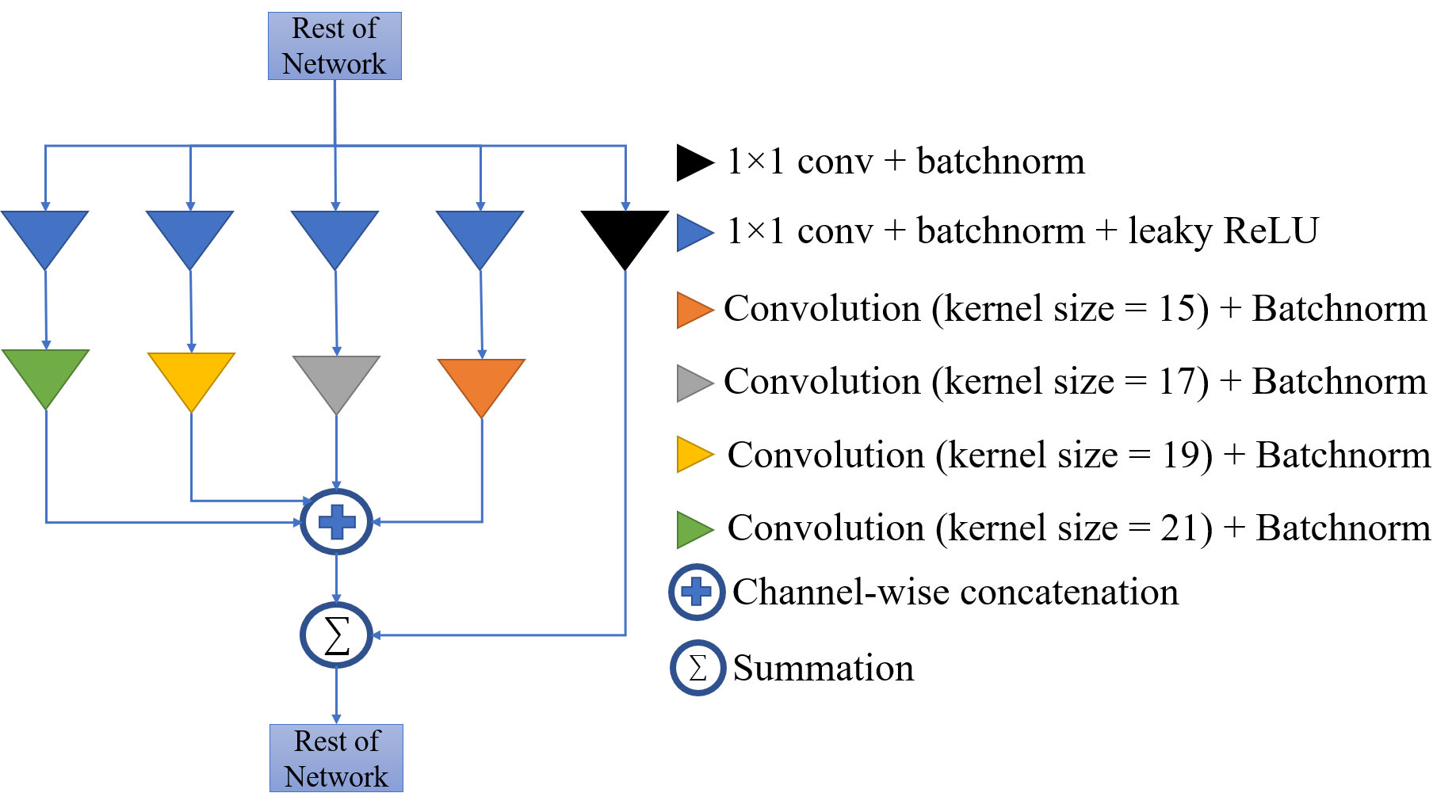}
    \caption{Inception Residual Block}
    \label{incres}
\end{figure}
\subsection{Naive CNN}
\label{naive-cnn}
The Naive CNN maps each window of the speech signal directly to its corresponding class. The components of the architecture are as follows:

\begin{enumerate}
    
    \item \textbf{Convolutional Block}: Each Convolutional Block consists of a Convolutional Layer\footnote{All convolutions refer to one dimensional convolutions unless specified.} of kernel size $k$, followed by Batch Normalization and ReLU activation, succeeded by another Convolutional Layer (kernel size = $k$), Batch Normalization and ReLU activation. The first convolution doubles the number of features and maintains the same length of the input feature map while the second convolutional layer maintains the same number of features and divides the length by half. Effectively, for an input feature map of dimension $(F,L)$ the dimension of the output feature map of the block is $(2 \times F, L/2 )$.   
    
    \item \textbf{Fully Connected Block}: This block consists of a series of Fully Connected Layers, each being succeeded by ReLU activation and a dropout of probability 0.5 to prevent overfitting. Depending on the complexity of the task, the number of activation units and number of layers is empirically altered. 

\end{enumerate}
The Naive CNN comprises of a $N$ Convolutional Blocks followed by a Fully Connected Block. The output feature map of the $n^{th}$  Convolutional Block has dimensions $(2^{n-1} \times c\ , l/2^n)$ where $c$ is the number of output channels of the first Convolutional Block and $l$ is the length of the input. The last Convolutional Block's output feature map is flattened before passing it on to the Fully Connected Block. Softmax activation is applied to the output of the last layer of the Fully Connected Block.

\begin{table*}

\centering
\renewcommand{\arraystretch}{1.3}
\caption{Model performance}
\begin{tabular}{|c|c|c|c|c|}
\hline
\textbf{Dataset}                  & \multicolumn{1}{c|}{\textbf{Model}}  & \multicolumn{1}{c|}{\textbf{Accuracy}} & \multicolumn{1}{c|}{\textbf{F1 Score}} & \multicolumn{1}{c|}{\textbf{Model Parameters}} \\ \hline
\multirow{2}{*}{EmoDB}   & \multicolumn{1}{c|}{Naive CNN} & \multicolumn{1}{c|}{0.843}    & \multicolumn{1}{c|}{0.829}    & \multicolumn{1}{c|}{172,368,455}        \\ \cline{2-5} 
                         & \multicolumn{1}{c|}{WaDeNet}   & \multicolumn{1}{c|}{0.934}    & \multicolumn{1}{c|}{0.928}    & \multicolumn{1}{c|}{\hphantom{M}47,156,391}         \\ \hline
\multirow{2}{*}{RAVDESS} & Naive CNN                      &        0.497                       &     0.484                          &       173,955,656                                \\ \cline{2-5} 
                         & WaDeNet                        & 0.872                         & 0.871                         & \hphantom{M}47,156,904                              \\ \hline
\multirow{2}{*}{VOICED}  & Naive CNN                      & 0.961                         & 0.947                         & \hphantom{M}90,057,282                              \\ \cline{2-5} 
                         & WaDeNet                        & 0.998                         & 0.997                         & \hphantom{M}26,182,306                              \\ \hline
\multirow{2}{*}{TESS}    & Naive CNN                      &          0.961                     &    0.962                           &                             173,953,607          \\ \cline{2-5} 
                         & WaDeNet                        & 0.982                         & 0.982                         & \hphantom{M}47,156,391                              \\ \hline
\end{tabular}
\label{NaivevsWade}
\end{table*}

\subsection{Wavelet Decomposition Net (WaDeNet)}
From the analysis of the architecture of the Naive CNN it is clear that there are some shortcomings. Firstly, increasing the depth of this network for better information capture will possibly introduce the problem of vanishing gradients. Secondly, the frequency composition of each window, which is vital for speech processing, is unknown to the model. 

In an effort to address these drawbacks, we propose \textit{Wavelet Decomposition Net  (WaDeNet)}, an enhancement of the Naive CNN. By inserting an Inception-Residual Block at the end of each Convolutional Block, WaDeNet has an increased network depth while alleviating vanishing gradients. The Inception-Residual Block consists of multiple convolutions with different kernel sizes performed in parallel as shown in Fig. \ref{incres}. Following this, channel wise concatenation of these representations is done. 

The second shortcoming is countered by introducing frequency information into the model by incorporating a multiresolutional representation. This is achieved by embedding the Wavelet Transform into the architecture, which is well experimented with for image reconstruction \cite{liu2019multi,fujieda2018wavelet,ramanarayanan2020dc}. As shown in Fig \ref{arch}, a representation of a particular order of the Discrete Wavelet Transformation (DWT) of the signal is concatenated with the output feature map of the preceeding Convolutional Block ($n$) and passed on to the succeeding Convolutional Block ($n+1$). 
The Haar wavelet was used for the decomposition of the signal. Each decomposition is passed to a $DWT\ Gate$ which consists of a series of Convolutions that produces the representation.
Thus for a total of $N$ Convolutional Blocks there are $N$ levels of wavelet decomposition. 

This approach is fundamentally different from using signal processing techniques to extract features that consist of both time and frequency information followed by training a classifier using these features. By incorporating the DWT into the network, the temporal information of the signal is not only retained but also learnt on in an end-to-end manner. Additionally, through feature map concatenation, the number of parameters is drastically reduced while improving performance, as discussed in Section \ref{Discussion}. This makes it suitable for mobile health applications such has emotion and stress monitoring. 

\section{Experiments and Results}


\subsection{Dataset and Preproccesing}
\label{dataset}
We run our experiments on popular datasets which consist of speech data suitable for mobile health applications such as emotion recognition and pathological voice recognition. The datasets and their respective descriptions are as follows:

\begin{enumerate}
    \item \textbf{EmoDB \cite{burkhardt2005database}:} This database consists of 10 actors who simulate 7 different emotions (anger, disgust, fear, happiness, pleasant surprise, sadness, and neutral) by uttering sentences used in everyday communication. There are 535 utterances in the dataset, amounting to 1487 seconds and an average utterance length of about 2.77 seconds. 
    \item \textbf{RAVDESS \cite{livingstone2018ryerson}: }The Ryerson Audio-Visual Database of Emotional Speech and Song (RAVDESS) consists of 12 male and 12 female actors speaking each expression in a North American Accent. The actors speak each statement with 8 different emotions - anger, calm, disgust, fear, happiness, sadness, surprise, neutral - at two levels of intensities, except when speaking in a neutral manner. 
    \item \textbf{VOICED \cite{cesari2018new}: }The VOice ICar fEDerico II Database (VOICED) contains 208 voice samples of adults between the age 18 and 70, of which 150 are pathological and the rest healthy. Each sample is consists of uninterrupted vocalization of the vowel 'a' for 5 seconds.  
    \item \textbf{TESS \cite{E8H2MF_2020}: }The Toronto Emotional Speech Set (TESS) consists of 200 words spoken by two female actors. Each of the voice samples emulate 7 different emotions namely anger, disgust, fear, happiness, pleasant surprise, sadness, and neutral. 
    
\end{enumerate}
The voice samples are segmented into windows of 320 ms each with a 75\% overlap. Each window is labelled depending on the dataset. For instance, a window in the EmoDB dataset is labelled with the corresponding emotion.    
\subsection{Implementation}
For our experiments, $N=4$, $c_i=64$ and $k=3$ for each convolution was chosen. Stochastic Gradient Descent with an initial learning rate of 0.001 was used to optimize the weights of the model by minimizing the categorical cross entropy. WaDeNet was trained for 150 epochs on 60\% of the data with a learning rate scheduler which reduced the learning rate by a factor of 10 after 50 epochs. These constants were empirically decided after extensive experimentation.
\subsection{Baseline Comparison}
\label{Discussion}
We compare the performance of the Naive CNN and WadeNet on the datasets discussed in the previous section. Since the F1 score is a more reliable metric for multi-class classification problems, we report it in addition to the overall accuracy of the model's performance on the test data (20\% of the total data). 
 
It is evident that the absence of spectral features translates to a subpar performance as evident from the performance of the Naive CNN (Table \ref{NaivevsWade}). In contrast, WaDeNet outperforms the Naive CNN with an average increase in accuracy and F1 score of 13.1\% and 13.9\% respectively. Additionally, WaDeNet has a maximum improvement up to 38.7\% in the F1 score on the RAVDESS dataset. 

Furthermore, WaDeNet is considerably lighter than the Naive CNN, with an average reduction of 72.26\% in the number of model parameters. 
\subsection{Benchmarking}
\label{benchmark}
In order to validate WaDeNet, we evaluate WaDeNet's performance against existing research work that portray state of the art performance on the datasets considered. The experimental conditions for each benchmark were replicated as specified in the respective work. As seen in Table \ref{FutureWorks}, WaDeNet performs considerably better than the current state of the art models across all the datasets, with an average increase in accuracy of 6.36\%, with a maximum of 11.4\% on the RAVDESS dataset.

\begin{table}
\renewcommand{\arraystretch}{1.3}
\caption{Benchmarking Performance}
\label{FutureWorks}

\centering
\begin{tabular}{|c|c|c|c|}
\hline
\multirow{2}{*}{\textbf{Dataset}} &

 \textbf{Number}  &

 \multicolumn{2}{c|}{\textbf{Accuracy}} \\ \cline{3-4} 
                      & \textbf{of classes} & \textbf{Benchmark}      & \textbf{WaDeNet}      \\ \hline
EmoDB  & 7 & 0.870 \cite{benchmark} & 0.934\\
\hline
RAVDESS  & 8 & 0.757 \cite{bhavan2019bagged} & 0.871\\
\hline
VOICED                    & 2 &  0.984 \cite{chen2020deep}         & 0.997        \\ \hline
TESS  & 7 & N/A & 0.982 \\ \hline
\end{tabular}
\end{table}


\section{Conclusion and Future Work}
In this paper we propose WaDeNet, an end-to-end architecture capable of learning spectral information through Wavelet Decomposition from a raw speech signal. WaDeNet outperforms the Naive CNN with the same number of fully connected layers with a 72.26\% reduction in model parameters. Moreover, WaDeNet also beats the current state of the art accuracy on all the datasets considered in Section \ref{dataset}. Since WaDeNet is end-to-end, it does not require disjoint training processes of different modules. This advantage combined with the benefits of a light model makes it suitable for mobile speech processing applications.

Additionally, we intend to perform quantization and replace standard convolutions with mobile convolutions to further optimize the model for mobile deployment.
\bibliographystyle{IEEEbib}
\bibliography{citations}

\end{document}